\documentstyle[12pt]{article}
\renewcommand{\baselinestretch}{1.3}

\begin{document}

\newfont{\abg}{cmsl12}  \def\otrue{{\bf{\omega}}}  
\def\bfB{\mbox{{\abg {\bf B}}}}
 \def\gtrue{\mbox{{\abg {\bf g}}}}\def\bfg{\gtrue}
\pagestyle{myheadings}                                        
\newcommand{\pref}[1]{\ref{#1}\fbox{#1}}                      
\newcommand{\plabel}{\label}
\newcommand{\prefeq}[1]{Gl.~(\ref{#1}\fbox{#1})}              
\newcommand{\prefb}[1]{(\ref{#1}\fbox{#1})}                   
\newcommand{\prefapp}[1]{Appendix~\ref{#1}\fbox{#1}}          
\newcommand{\plititem}[1]{\begin{zitat}{#1} {\fbox{#1}}a
                                             \end{zitat}}     
\newcommand{\plookup}[1]{\hoch{\ref{#1}}\fbox{#1} }           
\newcommand{\pcite}[1]{\cite{#1}\fbox{#1} }                  
\newcommand{\hinweis}[1]{\fbox{#1}}                           


\pagestyle{myheadings}

\let\oe=o
\def\Di{\displaystyle}
\def\nn{\nonumber \\}
\def\sr{\stackrel}
\def\be{\begin{equation}}
\def\ee{\end{equation}}
\def\ba{\begin{eqnarray}}
\def\ea{\end{eqnarray}}
\def\pl{\plabel} 
\def\re{(\ref }
\def\rz#1 {(\ref{#1}) }   \def\ry#1 {(\ref{#1})}
\def\el#1 {\plabel{#1}\end{equation}}

\let\a=\alpha \let\b=\beta \let\g=\gamma \let\d=\delta
\let\e=\varepsilon \let\ep=\epsilon \let\z=\zeta \let\h=\eta \let\th=\theta
\let\dh=\vartheta \let\k=\kappa \let\l=\lambda \let\m=\mu
\let\n=\nu \let\x=\xi \let\p=\pi \let\r=\rho \let\s=\sigma
\let\t=\tau \let\o=\omega \let\c=\chi \let\ps=\psi
\let\ph=\varphi \let\Ph=\phi \let\PH=\Phi \let\Ps=\Psi
\let\O=\Omega \let\S=\Sigma \let\P=\Pi
\let\Th=\Theta \let\L=\Lambda \let \G=\Gamma \let\D=\Delta
\def\q{\quad} \def\qq{\qquad}
\def\w{\wedge}
\def\0{\over} \def\1{\vec} \def\2{{1\over2}} \def\4{{1\over4}}
\def\5{\bar} \def\6{\partial}

\def\({\left(} \def\){\right)} \def\<{\langle} \def\>{\rangle}
\def\lb{\left\{} \def\rb{\right\}}
\def\[{\lbrack} \def\]{\rbrack}
\let\lra=\leftrightarrow \let\LRA=\Leftrightarrow
\def\ul{\underline}
\def\wt{\widetilde}
\let\Ra=\Rightarrow \let\ra=\rightarrow
\let\la=\leftarrow \let\La=\Leftarrow

\let\ap=\approx \let\eq=\equiv \let\hc=\dagger
\let\ti=\tilde \let\bl=\biggl \let\br=\biggr
\def\CL{{\cal L}} \def\CX{{\cal X}} \def\CA{{\cal A}} \def\CE{{\cal E}}
\def\CF{{\cal F}} \def\CD{{\cal D}} \def\rd{\rm d}
\def\rD{\rm D} \def\CH{\cal H} \def\CT{{\cal T}} \def\CM{{\cal M}}
\def\CI{{\cal I}} \newcommand{\dR}{\mbox{{\sl I \hspace{-0.8em} R}}}
\newcommand{\dN}{\mbox{{\sl I \hspace{-0.8em} N}}}
\def\CP{{\cal P}} \def\CS{{\cal S}} \def\C{{\cal C}}
\def\CR{{\cal R}}

\def\bfPs{{\bf{\Ps}}}  
\def\Dsl{\not{\!\! D}} 
\def\dsl{\not{\! \partial}}
\def\tDsl{\not{\!\! \widetilde{D}}}
\def\bfe{{\bf{e}}}
\def\dlr{\stackrel{\lra}{\6}}
\def\dl{\overleftarrow{\6}}
\def\dr{\overrightarrow{\6}}
 
\begin{titlepage}
\renewcommand{\thefootnote}{\fnsymbol{footnote}}
\renewcommand{\baselinestretch}{1.3}
\begin{center} 
{\Large{Generalized 2d dilaton gravity with matter fields}}\\ 
\vspace{2cm} 
HEIKO PELZER\footnote{e-mail: pelzer@ghi.rwth-aachen.de}\\
THOMAS STROBL\footnote{e-mail: tstrobl@physik.rwth-aachen.de}\\
Institut f\"ur Theoretische Physik\\
RWTH-Aachen\\
D52056 Aachen\\
Germany\\
\medskip
\begin{abstract}
  We extend the classical integrability of the CGHS model of 2d
  gravity [1] to a larger class of models, allowing the gravitational
  part of the action to depend more generally on the dilaton field
  and, simultaneously, adding fermion-- and $U(1)$--gauge--fields to the
  scalar matter. On the other hand we provide the complete solution of
  the {\em most}\/ general dilaton--dependent 2d gravity action coupled
  to {\em chiral}\/ fermions. The latter analysis is generalized to a
  chiral fermion multiplet with a non--abelian gauge symmetry as well
  as to the (anti--)self--dual sector $df = \pm \ast df$ of a scalar
  field $f$.
\end{abstract}
\end{center}
\end{titlepage}
\setcounter{footnote}{0}
\section{The Setting and Our Results}
One of the most influential papers written on low-dimensional gravity
models was the one of Callan, Giddings, Harvey and Strominger
\cite{CGHS}. Motivated by studying Hawking radiation, they proved the
complete classical solvability of string inspired dilaton gravity
\cite{dil} coupled minimally to massless scalar fields.  Here we study
a similar problem: First we want to see how the analysis changes, if
the dilaton part of the action is replaced by a more general one.
Secondly, we want to advocate the study of the coupling to 
fermion and gauge fields.

The most general 2d gravity action for a metric $\gtrue$ and a dilaton
field $\Phi$ which yields second order differential equations is of
the form \cite{Banks}: \be I_{gdil}[ \gtrue ,\Phi]= -\2 \int_{\CM} d^2x
\sqrt{-\det \gtrue} \; [U(\Phi)\CR(\gtrue)+ V(\Phi) \gtrue^{\m\n}
\6_{\m}\Phi \6_{\n}\Phi +W(\Phi) ] \, .  \el gdil Here $\CR(\gtrue)$
denotes the Ricci scalar of the Levi-Civita connection of $\gtrue$ and
$U$, $V$, and $W$ are some arbitrary (reasonable) functions
 (``potentials'') of the dilaton.  Eq.\ \rz gdil reduces to the original
string inspired dilaton action upon the choice
$U(\Phi)=\exp(-2\Phi)/\pi$, $V=-4U$, and $W = -4 \l^2 U$.\footnote{In
  the notation of \cite{CGHS}, with the other sign convention for
  $\CR$.}

Generalizing the standard actions for real massless scalar fields $f$ and
fermionic fields ${\bf{\Ps}}$ in curved spacetime \cite{Birrel} by
allowing for dilaton dependent couplings $\beta(\Phi)$ and
$\gamma(\Phi)$, one has: \ba I_{scal}&=&\2 \int d^2x \,
\beta(\Phi) \, \sqrt{-\det \gtrue} \; \gtrue^{\m\n}
\6_{\m}f \6_{\n}f \, , \plabel{scal} \\
I_{ferm}[{\bf e},{\bf{\Ps}}]&=&\2 \int_{\CM} d^2x \, \gamma(\Phi) \, \det
({\bf e}_\r^b) \, \lb i \( {\overline{{\bf{\Ps}}}} \s^{a}{\bf e}_{a}^{\m }
D_{\m } {\bf{\Ps}} \) + \mbox{herm.\ conj.} \rb \, .  \plabel{ferm1}
\ea The indices $\m$ and $a$ take values in $0,1$ and $+,-$,
respectively. ${\bf e}_{a}^{ \m }$ is the inverse of ${\bf e}_\m^a$,
the component-matrix of the zweibein ${\bf e}^\pm={\bf e}^\pm_\m
dx^\m$, related to $\gtrue$ as usual: \be \gtrue = 2{\bf e}^+{\bf e}^-
\equiv {\bf e}^+ \otimes {\bf e}^- + {\bf e}^- \otimes {\bf e}^+ \, .
\el E $D_\m$ denotes the covariant derivative: $D_\m = \6_\m + \2
\varpi_\m \s^3$, 
where $\varpi_\m$ is the (torsion-free) spin connection. (In the
presence of an additional $U(1)$ gauge field $\cal A$, $D_\m$ is
understood to contain also the standard ${\cal A}_\m$ part, cf.\ 
below). The basic elements of the Clifford algebra have been
represented by Pauli matrices $\vec \s$, furthermore, with $\s^\pm
\equiv \2(\s^1 \pm i \s^2)$ and $\overline{{\bf{\Ps}}} \equiv
\bf{\Ps}^\dagger \s^1$, where ${\bf{\Ps}}$ is a two-component complex
column vector (the entries of which may be taken anti--commuting or
commuting, as one prefers, since we will stay on the classical level 
throughout this paper).

The global $U(1)$-symmetry of $I_{ferm}$ may be turned into a local
one by the standard procedure: $D_\m \rightarrow D_\m + i\, {\cal
  A}_\m$, where $\cal A$ is an (abelian) connection one-form. Dynamics
for $\cal A$ is generated by \be I_{U(1)} = \frac{1}{4} \int_{\CM} d^2x
\sqrt{-\det \gtrue} \; \alpha(\Phi) {\cal F}_{\m \n} {\cal F}^{\m \n}
\; , \pl{U1} \ee where ${\cal F} = d {\cal A}$ and $\alpha(\Phi)$ is
the again dilaton dependent coupling.

\vskip8mm

\centerline{\em The results and the organization of the paper are as follows:} 

\vskip2mm

In {\underline {Sec.\ 2}} we derive the general field equations of the
action \be I = I_{gdil} + I_{scal} + I_{ferm} + I_{U(1)} \, .
\plabel{I}\ee In {\underline {Sec.\ 3}} we provide the {\em general}\/
(local) solution to the coupled system \re{I}) restricted as
follows:\footnote{By ``general solution'' we always mean the following
in this paper: All solutions of the field equations of the respective
Lagrangian --- a coupled system of partial differential equations ---
without imposing any boundary conditions. Thereby we identify all
constants and functions parametrizing the general solution which are
related by gauge transformations ($U(1)$, diffeomorphism, or Lorentz,
also not restricted by any boundary or overlap conditions). Thus the
solutions are local solutions, solutions on trivial topology $\dR^2$,
which are also not yet geodesically complete or maximally extended in
general. A maximal extension of these solutions and the study of
solutions and their parameters on non--trivial topologies, as done in
\cite{PartII} for the case of $I_{gdil}$ alone, is not contained in
the present analysis.} The ``potential'' $W(\Phi)$ is determined by
the freely chosen functions $U(\Phi)$ and $V(\Phi)$ up to the choice
of two (real) constants $a$ and $b$ through the relation \be W(\Phi) =
\exp \( - \int^\Phi \frac{V(z)}{U'(z)} dz\) \, \left(4a \, U(\Phi) +
2b\right) \,\, , \plabel{relation} \ee $\beta(\Phi)$ is constant,
$\gamma(\Phi)$ arbitrary, and $\alpha(\Phi)$ is determined up to one
constant by a similar equation as the one for $W$ (Eq.\ \re{a1})
below), which includes $\a =0$ (no gauge fields).

The treatment in Sec.\ 3 constitutes a reasonable generalization of
the CGHS model. The latter satisfies Eq.\ (\ref{relation}) with $a=0$
and $b=-2\l^2$, the coupling to the scalar fields is minimal, $\beta
\equiv 1$, and in that model there are no fermions or gauge fields
($\gamma=\alpha=0$). The class of models considered in Sec.\ 3
contains also e.g. the one--parameter generalization of the CGHS model
considered in \cite{Fabbri} (with the {\em same}\/ $a$ and $b$ as
above). Similarly, we also generalize the recent (independent)
considerations of Cavagli\`a et al.\ \cite{Cava1}, where they consider
fermions coupled to generalized dilaton gravity; for $a=0$ they
provide the general solution
to the fermion--gravity system, while for the more complicated case
$a\neq 0$ they find the stationary solutions only. Some
further related and in part complementary work is \cite{Rest}.

Note that as we allow also for $U(1)$ gauge fields, the classically
solvable models considered in Sec.\ 3 incorporate generalizations of
the massless Schwinger model to a non--trivial gravitational sector.
For reasons of better interpretation, all the matter dependence of the
metric is expressed in terms of the energy momentum tensor in that
section, furthermore.

We do not know how to find the general solution of the field equations
of the total gravity--matter action 
$I$ for completely arbitrary choices of $U,V,W,\a,\b$, and $\g$.
However, as we will show in {\underline {Sec.\ 4}} and discuss briefly
below, the general system \re{I}) may be solved in the case of {\em
  chiral}\/ fermions (with our without $U(1)$ gauge fields) and no
scalar fields ($\b=0$).  This solution may be generalized,
furthermore, to the presence of non--abelian gauge fields with a
chiral fermion multiplet in the funamental representation.  Also we
may allow for additional torsion dependent terms in the 
gravitational part of the action. In Sec.\ 4 we thus generalize
the observation of W.\ Kummer \cite{Kummer} that the Katanaev--Volovich model
\cite{KV} of 2d gravity with torsion may be solved when coupled to
chiral fermions.

In {\underline {Sec.\ 5}} we will see that for  scalar fields coupled to
the gravity action $I_{gdil}$ there is an analog of the
above chiral fermion solutions.  These are the (anti-)selfdual
solutions $df=\pm \ast df$, where the star denotes the Hodge dual
(with respect to the dynamical metric). As we will show, such
solutions exist for constant coupling $\beta$ only. This excludes the
particularly interesting case of spherically reduced 4d scalar fields,
otherwise still incorporated in $I_{gdil}+I_{scal}$, since there
$\beta \propto \Phi^2$ (with an appropriate definition of
$\Phi$).  For a minimal coupling $\beta = const$, however, the general
(anti--)self --dual solutions may be written down explicitly. More
specifically, we will be able to find a mapping between the
self--dual sector of $I_{gdil} + I_{scal}$ and the chiral sector
$I_{gdil} + I_{ferm}$. The former will be seen to be a subsector of
the latter. In a way this might be called a classical version of
``bosonization''.

We finally remark that many of the results presented in this paper
have already been obtained in \cite{Diplom} as an outcome of joint
work.

\vskip8mm

\noindent Let us discuss the solutions in some further detail: 

\vskip2mm

If there are no matter fields present
\cite{Classgendil,PartI}, the metric $\gtrue$ always has
a Killing vector and thus may be brought into the form \be \gtrue = 2
dx^0 dx^1 + h_0(x^0) (dx^1)^2 \, \, . \label{EF} \ee For a given
Lagrangian (\ref{gdil}), there is a one--parameter family of functions
$h_0$, which may be written down explicitly in terms of the
``potentials'' $U,V$, and $W$.  The dilaton field $\Phi$, moreover,
also respects the Killing symmetry, being an explicitly known function
of $x^0$, determined by $U,V$, and $W$. In the case of pure gravity, 
the space of local solutions is one--dimensional (``generalized 
Birkhoff theorem'').

Maximal extension of the local solutions \re{EF}) generically leads to
a variety of spacetimes with black holes and kinks \cite{PartII,Kinks}. For
a given model $I_{gdil}$ with a {\em ``generic''}\/ 
choice of $U,V$, and $W$, there
are globally smooth solutions on two--surfaces of arbitrary
non--compact topology; without matter fields and for a fixed topology
of spacetime this solution space (space of global solutions to the
field equations modulo gauge transformations) is still finite
dimensional.

For the local solutions found in Sec.\ 4 (but without gauge fields) 
and those found in Sec.\ 5, 
Eq.\ (\ref{EF}) is modified into \be \gtrue = 2 dx^0 dx^1 + [h_0(x^0)
+ k_0(x^0) h_1(x^1)] \, (dx^1)^2 \, \, , \plabel{CEF} \ee where $h_0$
is the same function as the one in the pure gravity case, Eq.\ \re{EF})
above, and the dilaton field $\Phi=\Phi(x^0)$ remains unchanged,
too. The respective matter field depends on the null coordinate $x^1$
only.  The function $k_0$ in Eq.\ \re{CEF}) is determined completely by
$U,V$ and $W$. Thus all the matter dependence in \re{CEF}) may be put
into the function $h_1$. More explictly we will find \be h_1 =
-2\int^{x^1} {\bf T}_{11}(u)du \, , \pl{h1} \ee where ${\bf T}_{11}={\bf T}_{11}(x^1)$
is the only non--vanishing component of the energy momentum tensor of
the respective matter field. The space of local solutions is infinite
dimensional now, being parametrized by the initial data for the matter
field and the one constant in $h_0$. Moreover, by means of matter
fields one may generate transitions between various sectors of the
{\em pure}\/ gravity solutions (beside the generation of completely new, not
yet investigated sectors, certainly); in several instances this will
also include the possibilty of a black hole formation due to the
presence of matter fields (but cf.\ also the remarks on the missing
``Choptuik effect'' following below).

Allowing for a non-vanishing $U(1)$ gauge field in Sec.\ 4 there
is an additional contribution $K_0(x^0) E^2(x^1) \,
(dx^1)^2$ to $\gtrue$, where $K_0$ is again determined by means of
$U,V,W$ and $E$ is the electric field of $\cal A$: $E\equiv -\alpha
\ast dA$; up to a constant of integration, $E(x^1)$ is determined by
the initial data of the fermion field.  An analogous form of the
solution holds for $n$ chiral fermion generations with e.g. $SU(n)$
gauge symmetry; $E^2(x^1)$ is then just replaced by $tr(E^2)(x^1)$. 

We finally remark that a ``superposition'' of the scenarios of Secs.\
4 and 5 is possible: The general solution of the chiral fermion {\em
and}\/ (anti--)selfdual scalar field sector of the {\em general}\/
system \re{I}) (or any non--abelian generalization thereof) may be put
together easily from the formulas of those sections.

The metric of the solutions found in Sec.\ 3 may be brought into the
form \be \gtrue = \frac{4(2a\phi +b) \, dx^+\,dx^-}{(1+ax ^+x^-)^2 \,
  W(U^{-1}(\phi))} \, , \label{hihi} \ee where the function $\phi$
contains all the dependence on the matter fields.  In terms of the
total energy momentum tensor ${\bf T}_{\m\n}$ of the matter fields it
has the form given in Eq.\ \re{phiU1}) below. In the case of no gauge
fields $(\a =0)$ $\phi$ simplifies somewhat, since then ${\bf T}_{+-}=0$ and ${\bf
  T}_{\pm\pm} ={\bf T}_{\pm\pm}(x^\pm)$.  For the CGHS model,
furthermore, in addition 
$W(U^{-1}(\phi))=-4\l^2 \, \phi$, $a=0$ and $b=-2\l^2$,
so that the conformal factor in Eq.\ \re{hihi}) reduces to $1/\phi$
and Eqs.\ \re{phiU1}) may be seen to reproduce the
standard result $\;\,$ $\phi=\int dx^+ \int dx^+ \, {\bf T}_{++}(x^+) +\int
dx^- \int dx^- \, {\bf T}_{--}(x^-) + 2\l^2x ^+x^-$.

The latter solution retains its form when fermionic matter is added to
the action of the CGHS model or when fermionic matter replaces the
scalar one.\footnote{Note that this is no more true, when we also turn
  on a $U(1)$ gauge field; ${\bf T}_{++}$, e.g., depends on both
  coordinates $x^\pm$ then and $\phi$ contains also ${\bf T}_{+-}$.}
It appears to us that fermionic matter may have its advantages over
scalar matter: An energy momentum shock wave, e.g., may be generated
simply by a discontinuity in the phase of the fermions (as opposed to
scalar fields where one needs the ``square root'' of a delta
function). More important, there is no problem in quantizing 2d
massless fermions, while quantized massless scalar fields, strictly
speaking, do not exist in two dimensions (cf.\ \cite{massless2d}).

The metric \re{hihi}) may be expected to describe interesting
generalizations of the CGHS black holes. We will not pursue such
global aspects in this paper, however. 

Upon restriction to the chiral and selfdual sector of the fermion and
scalar fields, respectively, the solution \re{hihi}) has to be of the
form \re{CEF}) up to a change of coordinates. We will also not
construct the necessary diffeomorphism explicitly, although this might
provide some further insights into the physics of the solutions.

Despite the many physically attractive features of the CGHS model,
such as the possibility to discuss Hawking radiation in a classically
soluable model, there is also one qualitative difference to the
spherically symmetric sector of 4d Einstein gravity with scalar
fields, which may be described by $I_{gdil} + I_{scal}$ for some other
specific choice of the potentials $U,V,W$ and $\beta$: In the Einstein theory
the density of matter has to surpass a threshold before a black hole
can be formed by the collapse of matter; below this threshold the
matter will just scatter apart again (cf., e.g., the numerical work of
Choptuik \cite{Chop}).  In the CGHS model, on the other hand, already
the smallest possible contribution to the energy momentum tensor of a
scalar field will generate a black hole from the Minkowski
vacuum.\footnote{We are grateful to K.\ Kucha\v{r} for pointing this
  out to us.}

We did not check, if this undesirable feature is shared by all the
models considered in Sec.\ 3. The solutions found in the subsequent
Secs.\ 4 and 5 do miss this Choptuik effect, however, at least if we
are interested in true curvature singularities when speaking of black
holes.\footnote{In the context of 2+1 dimensional gravity people speak
  of black holes also for globally smooth constant curvature solutions
  where a region containing closed timelike curves is protected from
  the ``outside'' world by an event horizon \cite{BTHZ}.} This may be
seen as follows: The scalar curvature of a metric of the form
(\ref{CEF}) is \be {\cal R} = - {\ddot h_0}(x^0) - h_1(x^1) \, {\ddot
  k_0}(x^0) \, .\label{uuu}\ee Suppose the matter--free solution with
$h_0$ describes a flat or at least non--singular spacetime for some
choice of the parameter in the vacuum solution. Then a curvature
singularity can result only from the divergence of the second term in
\re{uuu}) for some $x^\m$. However, a smooth choice of initial data
for the matter fields yields a smooth function $h_1$.  Thus a
divergence of $\cal R$ can result only when $\ddot k_0(x^0)$ blows up
for some value of $x^0$. As a consequence of the explicit form of
$h_1$, Eq.\ \re{h1}), there is again no threshold for a curvature
singularity to form: if $k_0$ gives rise to such a singularity, this
singularity will show up {\em whenever}\/ $h_1$ is just non zero.  We
suspect that for the system \re{I}) a Choptuik effect will be present
only when the scalar fields are coupled non--minimally, i.e.\ when
$\beta \neq const$.

Our investigation of chiral fermions was inspired by the observation
of W.\ Kummer \cite{Kummer} that, when coupled to the KV model of 2d
gravity with torsion \cite{KV}, chiral fermions allow for a general
solution (cf.\ also \cite{Solodukhin} for an independent, but later
work with the same conclusion). Our work generalizes this result to a
broad class of models of 2d gravity as well as to the case of several
fermion generations with an additional gauge symmetry.

Moreover, we think that our considerations in Sec.\ 4 also simplify
and illuminate the analysis of \cite{Kummer,Solodukhin}: As shown in
previous work \cite{PartI}, the purely gravitational part $I_{gdil}$
of the action may be reformulated in terms of a Poisson $\s$--model
\cite{PS}, the action of which has the form \ba L(A_i,X^{i})=\int_{\CM}
A_{i} \w dX^{i} + \2 \CP^{ij}(X(x))A_{i} \w A_{j}\q , \plabel{PS} \ea
where ${\cal P}^{ij}$ is determined explicitly through $U,V,W$, the
indices $i,j$ run over three values, and the $A_i$ and $X^i$ are three
one--froms and three zero--forms, respectively (further details, also
about subsequent statements, will be provided in Sec.\ 4 below).  It
turns out that the addition of $I_{ferm}$ to the gravi\-tational action
gives rise to a simple modification of \re{PS}): We merely have
to add a term \be \int A_i \wedge J^i \label{min} \, , \pl{AJ} \ee
where the one--forms $J^i$ contain fermion variables only. Now one of
the main ideas in the realm of $\s$--models is the use of advantageous
coordinates on the {\em target}\/ space of the theory, which in \re{PS}) are
the $X^i$. As demonstrated in \cite{PS} and more explicitly in
\cite{PartI}, it is possible to choose coordinates $\wt X^i(X)$ such
that the tensor ${\cal P}^{\tilde i\tilde j}(\wt X)$ takes a simple
form, which, in particular, does no more depend on the (new) field 
variables $\wt X^i$.
Now, in the presence of an additional term \re{min}), such a procedure
will be helpful only if the transformation does not induce too
complicated an $\wt X$--dependence of the transformed current $J^{\wt
  i}= \left(\partial\wt X^i/\partial X^j\right) \, J^j$. 
So, in the combined system the task is to find target space
coordinates such that the two structures ${\cal P}$ and $J$ {\em
  simultanously}\/ take a simple form. As we will see, this is most
straightforward if the fermions are chiral. For both chiralities we
did not succeed to simplify the coupled system sufficiently.

The above discussion generalizes to a multiplet of fermions with the
additional presence of a non--abelian gauge symmetry; the full
Lagrangian $I_{gdil}+I_{ferm}+I_{YM}$ then again takes the form of Eqs.
\re{PS},\ref{AJ}), where now $\CP^{ij}$ depends also on the YM--coupling
$\a(\Phi)$ and the indices $i,j$ run over a larger set of values.
With some adaptation the above discussion applies also to the
(anti--)self--dual sector of $I_{gdil}+I_{scal}$, as discussed in Sec.\ 5. 

The analysis within this paper remains on the purely classical level.
Much (although not all) of the interest in two--dimensional models
results from their
quantum aspects. We do not know, if the classical solutions
found in this paper survive quantization. The chiral solutions could
be plagued by chiral anomalies and the quantum integrability of the
models in Sec.\ 3 by anomalies of the diffeomorphism constraints. For
a somewhat controversial discussion of the latter topic in the case of
the CGHS model cf.\ \cite{JackiwKuchar}; for another attempt to
quantize the CGHS model see \cite{Mikovic}, for semiclassical aspects
we refer to the original work \cite{CGHS} as well as to
\cite{Strominger} (and references therein). For the quantization of
the purely gravitational action \re{gdil}) cf.\ 
\cite{KunstatterStrobl,PS}; quantization of models resulting from
specifications of $U,V,W$ in \re{gdil}) may be found also in the papers
\cite{Kastrupetc}. For thermodynamics of 2d black holes we refer
the reader to \cite{Lau} and \cite{Wald}, the latter method being
applicable to theories in any dimension $d \ge 2$.

\section{Field Equations of the General Model}
The variation of $I_{scal}$ and $I_{ferm}$ with respect to its matter
content yields the following generalizations of the massless
Klein--Gordon and Dirac equation: \ba \Box f + \(\ln |\b |\)' \6^\m
\Phi \6_\m f &=& 0 \pl{KG}\\ \tDsl \bfPs &=& 0 \pl{Dirac} \ea where
$\Box = \nabla^\m \nabla_\m$, with $\nabla$ the Riemannian covariant
derivative, and where prime denotes differentiation with respect to the
argument of the respective function, a convention kept throughout this
paper. Furthermore,
$$\tDsl = \bfe_a^\m \s^a \wt{D}_\m \, , \quad \wt{D}_\mu = \6_\m - \2
\s^3 {\bf{\o}}_\m + i \wt{\CA}_\m \, , \quad \wt{\CA}_\m \equiv
{\CA}_\m -i \6_\m \ln \sqrt{|\gamma(\Phi)|} \, .  $$
So both equations
\re{KG}) and \re{Dirac}) are modified by a term resulting from the
dilaton dependent coupling. For the fermions, however, the
modification has a particularly simple form: One merely has to add an
imaginary part to the $U(1)$ connection $\CA$. Moreover, since this
imaginary part is an exact form, the modification may equally well be
absorbed into a fermion field with a redefined absolute value:
$\wt{\bfPs} := \sqrt{|\gamma(\Phi)|} \bfPs$ satisfies the standard
Dirac equation in curved spacetime, $\Dsl \wt{\bfPs} = 0$. This
becomes also obvious from the form of the action \re{ferm1}), since an
unwanted derivative from the first part of the action is seen to
cancel against the respective hermitean conjugated term.

This last observation establishes also the conformal invariance of the
fermionic action. While the action for the scalar field $f$ is
conformally invariant with unmodified $f$ ($f$ has conformal weight
zero), $\bfPs$ transforms with the inverse fourth root of the
conformal factor ($\bfPs$ has conformal weight minus one half). Still
the total action is not invariant under conformal (or Weyl)
transformations due to the presence of the gravitational and $U(1)$
part of the action.\footnote{The action for gauge fields is
conformally invariant only in four spacetime dimensions (and
  this on the classical level only); the trace of the energy momentum
  tensor, Eq.\ \re{TU1}) below, vanishes merely for $d=4$.}

Variation of the action $I$, Eq.\ \re{I}), with respect to the gauge
field $\CA$ yields: \be \nabla_\n \[ \a(\Phi) \CF^{\m\n} \] = \g(\Phi)
{\overline{{\bf{\Ps}}}} \bfe_a^\m \s^{a} {\bf{\Ps}} \pl{eomA} \, . \ee
Note that if an action contains a gauge field $\CA$, the solutions of
the coupled system do {\em not}\/ contain those of the system without
gauge field as a subsector: if $\CA \equiv 0$, the above equation
implies $\bfPs \equiv 0$. It is an additional equation without
counterpart in the system without a gauge field.


We now come to the variation of $I$ with respect to the metric
$\bfg_{\m\n}$ (or the zweibein $\bfe_\m^a$).  One finds \be \(
\nabla_\m \nabla_\n - {\bfg}_{\m\n} \Box \) \, U(\Phi) - \2
\bfg_{\m\n} \, \( V(\Phi) (\nabla \Phi)^2 + W(\Phi) \) + V(\Phi) \6_\m
\Phi \6_\n \Phi = {\bf T}_{\m\n} \, . \pl{eomg1} \ee Here ${\bf T}_{\m\n}$ is the
energy momentum tensor of the matter fields, ${\bf T}_{\m\n}=
{\bf T}^{scal}_{\m\n}+ {\bf T}^{ferm}_{\m\n} + {\bf T}^{U(1)}_{\m\n}$, where \ba
{\bf T}^{scal}_{\m\n} &=& \b (\Phi) \, \[\6_\m f \6_\n f - \2 \bfg_{\m\n}
\6_\m f \6^\m f \] \pl{Tscal} \, , \\ {\bf T}^{U(1)}_{\m\n} &=& \a(\Phi) \,
\[\CF_{\m \rho} \CF_{\n}{}^\rho - \frac{1}{4} \bfg_{\m\n} \CF_{\lambda
  \rho} \CF^{\lambda\rho} \] \pl{TU1} \, . \ea The energy momentum
tensor for the fermion fields is defined by ${\bf T}^{ferm}_{\m\n} = \bfe_{a
(\m} \, \(\d I_{ferm}/\d \bfe_{a}^{\n)}\) / \det ( \bfe_\rho^b)$,
where the brackets around the indices $\m$ and $\n$ indicate
symmetrization. For an action that depends on the vielbein only via
the combination $\gtrue_{\m\n}=\bfe_\m^a \bfe_{a\n}$ this reduces to
the standard expression ${\bf T}_{\m\n} = 2 \(\delta I_{scal}/\delta
\gtrue^{\m\n}\)/\sqrt{-\det{\gtrue})}$. (In such a case the
symmetrization in the indizes $\m$ and $\n$ would not be necessary,
furthermore).

The fermionic part of the action, Eq.\ \re{ferm1}), is of the form
$\int d^2x \, \det ( \bfe_\rho^b) \, \bfe_a^\m \, \bfB_\m^a$, 
resulting in ${\bf T}_{\m\n} = \bfe_{a(\m} \bfB_{\n)}^a - \bfg_{\m\n} \,
\bfe_{a}^{\rho} \bfB_{\rho}^a$ since $\bfB_\m^a$ does not depend on
the zweibein; this is a particular feature of two spacetime
dimensions, where the spin connection $\bf{\o}$ drops out from the
ferm\-ionic action altogether. As such ${\bf T}_{\m\n}$ is not tracefree.
However, it is straightforward to show that, as a consequence of the
field equations \re{Dirac}) with arbitrary $\gamma(\Phi)$,
$\bfe_{a}^{\rho} \bfB_{\rho}^a = 0$, so that ${\bf T}^{ferm}$ becomes
tracefree ``on--shell''. (We remark that the symmetrization in the
indices $\m$ and $\n$ is essential; the unsymmetrized version remains
non--symmetrical also on--shell).  The fermionic part of the energy
momentum tensor may now be written as \be {\bf T}^{ferm}_{\m\n} =
\bfe_{a(\m} \bfB_{\n)}^a \, , \quad \bfB_\n^a \equiv \gamma(\Phi) \,
\[{\overline{{\bf{\Ps}}}} \s^{a} 
(i \dlr_\n - \CA_\n) {\bf{\Ps}} \] \, ,\pl{Tferm} \ee
where $\dlr_\m \equiv (\dr_{\! \m} - \dl_{\! \m})/2$, with $\dl_{\!
  \m}$ acting on everything to its left except for $\gamma(\Phi)$ 
outside the brackets and $\dr_{\! \m} = \partial_\m$. 

The field equations \re{eomg1}) may be simplified by taking the trace
and eliminating $\Box U(\Phi)$:   \be \nabla_\m \6_\n U(\Phi)
+ \2 \bfg_{\m\n} \, W(\Phi) + V(\Phi) \( \6_\m \Phi \6_\n \Phi -\2
\bfg_{\m\n} \, \6_\m \Phi \6^\m \Phi \) ={\bf T}_{\m\n} -
\bfg_{\m\n} \, {\bf T}^{U(1)}  \, . \pl{eomg2} \ee 
Here ${\bf T}^{U(1)}$ denotes
the trace of the $U(1)$ part of the energy momentum tensor, the other
two contributions to the trace being zero as a consequence of the
conformal invariance of the respective parts of the action.

Finally, variation with respect to the dilaton $\Phi$ leads to the
equation 
\be - U' \, \CR + V' \, \6_\m \Phi \6^\m \Phi + 2V \, \Box \Phi - W' + 
\2 \a' \, \CF_{\m\n}\CF^{\m\n} + \b' \, \6^\m f \6_\m f + 2\gamma' 
\, {\bf e}_{a}^{\m } \frac{\bfB_\m^a}{\g}=0 \pl{eomPhi} \, ,\ee
with $\bfB_\m^a$ has been defined in Eq.\ \re{Tferm}). 

The action $I$ has several gauge symmetries: it is invariant
with respect to diffeomorphisms,  local Lorentz transformations in
the spin and frame bundle, and  $U(1)$ gauge transformations.
Despite the corresponding possible simplification of the field
equations, it is, to the best of our know\-ledge, beyond human abilities
to solve the general field equations following from $I$. But even in
the restricted cases discussed briefly in the Introduction and to more
detail in the following sections, we will not attempt to solve the
field equations directly. Instead we will make use of the covariance
of a Lagrange formulation with respect to a change of ``generalized''
coordinates. In the context of field theories generically local
transformations of field variables are preferred as usually (and
certainly also here) the action is local in the orignal field
variables. Such transformations will be used in all of the following
sections to simplify the system as much as possible already at the
level of the Lagrangian.

\section{General Solution for a Restricted Class of Lagrangians}
In this section we discuss the solutions to the Lagrangian $I$, Eq.\ 
\re{I}), with $\beta :\equiv 1$, $W$ subject to the condition
\re{relation}) for some constants $a$ and $b$, and $\alpha$ restricted
by a similar relation derived below in Eq.\ \re{a1}).


The matter part of the action (except for the gauge field part) is
conformally invariant (cf.\ also the preceding section). A conformal
transformation of the Ricci scalar, on the other hand, produces an
additive term with two derivatives on the conformal exponent (cf.,
e.g., \cite{WaldApp}).  Therefore it is near at hand to get rid of the
kinetic term for $\Phi$ in $I_{gdil}$, Eq.\ \re{gdil}), by a
$\Phi$-{\em dependent}\/ conformal transformation. This was first
proposed by H.\ Verlinde \cite{Verlinde} for string inspired dilaton
gravity and then generalized to
$I_{gdil}$ in \cite{BanksKunst}.  
We thus change
variables from $\gtrue$ to an ``auxiliary'' metric $g$ defined by:
\be g := \O^2(\Phi) \,\, \gtrue \, ,
\quad \O(u) \equiv \exp \[- \int^u \frac{V(z)}{2U'(z)} dz \] \, \pl{gneu} .
\ee Now the action $I_{gdil}$ as a functional of $g$ and $\Phi$ again
has the form of Eq.\ \re{gdil}), but with a potential $V$ that is
identically zero, while $U$ remains unchanged and $W$ is replaced by
$W(\Phi)/\O^2(\Phi)$. Due to the resulting 
absence of the $V$--term we now may use
$\phi := U(\Phi)$ instead of $\Phi$ to also trivialize the potential
$U$ to become the identity map.\footnote{This works locally only (and
  for a non--constant $U$ certainly, which we will assume); for
  the purpose of solving the field equations in local patches,
  this is sufficient --- except possibly for additional solutions
  where $\Phi$ takes the {\em constant}\/ value of an extremum of $U$
  and which would have to be analyzed separately. A similar remark may
  hold for the change of variables in \re{gneu}) certainly.} 
Distributing the conformal factor in Eq.\ \re{gneu}) equally on 
$\bfe^+$ and $\bfe^-$, $e^\pm := \O \, {\bf e}^\pm$, 
the spinors are transformed into $\hat{\psi}= \O^{-(1/2)} \bfPs$. 
It is, however, also possible to get rid of
$\gamma(\Phi)$ by means of the further redefinition (cf.\ the
discussion following Eq.\ \re{Dirac})): \be \psi :=
\sqrt{\frac{|\gamma(\Phi)|}{\O(\Phi)}} \,  \,\, \bfPs \, \pl{psi}.  \ee In
the new variables $g=2e^+e^-$, $\phi$, and $\psi$,  the
action $I$  
takes the form\footnote{Here and in what follows we assume $\gamma
  >0$; for $\g<0$, $B_\m^a$ is to be replaced by $-B_\m^a$ in Eq.\ 
  \re{J}).} \ba I[g,\phi,f,\psi,A]&= & \2 \int d^2x \, \sqrt{- \det g}
\; \big[ - \phi R(g) - \wt W(\phi) + \wt \b(\phi) \, \6^\m f \6_\m f +
\nn & & \mbox{}+ 2 e_a^{\m} B_\m^{a} + \2 \wt{\a}(\phi)
F^{\m\n}F_{\m\n} \big]
\, , \pl{J} \\
\mbox{with} && B_\m^{a} \equiv \overline{\psi} \s^{a} (i \! \dlr_\m -
\CA_\m) \psi \, , \quad \!
\wt W(U(\Phi)) =  \frac{W(\Phi)}{\O^2(\Phi)} \, ,\\
& & \wt{\a}(U(\Phi)) = \O^2(\Phi)\, \a(\Phi) \, , \quad \wt
\b(U(\Phi)) = \b(\Phi) \, \; . \pl{wtbeta} \ea  Indices are raised
by means of the ``auxiliary metric'' $g$ and we made use of $
\sqrt{- \det g} = \det e_\m^a$. Note that \re{J}) is in the same form
as \re{I}), just with the appropriate replacement of variables and a
specification of ``potentials'', resulting from putting some of the
general ones into the transformation formulas for variables.
Therefore the field equations in the new variables follow from those
of the previous section. 

The intention of the present section is to study models in which the
{\em auxiliary}\/ metric $g$ is one of constant curvature. Thus we
require that $\wt W(\phi)$ is at most linear in $\phi$, while $\wt \a$
and $\wt \b$ have to be independent of this field. The first condition
is precisely \re{relation}), $\wt W(\phi) = 4a \phi + 2b$ for two
constants $a$ and $b$, while the second one implies \be \a(\Phi) =
\a_0 \,
\exp \[\int^\Phi \frac{V(z)}{U'(z)} dz \] \pl{a1} \ee for some constant
$\a_0$;
$\b$, on the other hand, has to be constant altogether and we will
normalize it to $\b =1$. Note that, due to our definition of $\psi$,
we are not forced to also put $\g'$ to zero.

In the new variables the metric $g$ decouples completely from the
matter sector. Up to a choice of coordinates, $R(g)= -4 a$ forces $g$
to take the form \be g = \frac{2 dx^+ dx^-}{(1+a x^+x^-)^2} \equiv : 2
\exp (2\xi) dx^+ dx^- \pl{const} \, \, ,\ee where the function $\xi$
is defined by this equation. Note that this does not imply that the
gravitational and matter sectors of the original model decouple. The
situation is quite analogous to a system of coupled harmonic
oscillators: There the introduction of appropriate variables (``normal
coordinates'') leads to a system of decoupled harmonic oscillators.
Also here the field equations in the new variables $g$ etc.\ take
qualitatively the same form as the one of the original variables
$\gtrue$ etc., just that, upon restriction to the class of Lagrangians
considered in this section, in the new variables the equations of
motion simplify greatly and (in part) decouple.  (Certainly, if one
counts the dilaton to the gravitational sector, this decoupling is not
complete. The equations for the dilaton depend on the energy momentum
tensor, cf.\ Eqs.\ \re{eomg2}) or Eqs.\ \re{dil1},\ref{dil2}) below.
It is also in this way that the matter enters the metric $\gtrue$,
cf.\ Eq.\ \re{gneu}) above).

Due to $\b = \wt \b= 1$ we find $f$ to be a superposition of left--
and right--movers, \be f = f_+(x^+) + f_-(x^-) \pl{Lsgphi} \,\, , \ee
just as in flat Minkowski space. This is the case since due to the
conformal invariance of \re{KG}) with $\b' =0$ and $f$ having
conformal weight zero, $\Box f = 0$ reduces to just $\6_+ \6_- f=0$
for any metric in the conformal gauge.  So here $\gtrue \to g$ does
not make any difference.

\vskip5mm

{\centerline{\em No gauge fields ($\a_0 = 0$):}}

\vskip2mm

Next we turn to the field equations for the redefined fermion fields
$\psi$.  For simplicity we restrict ourselves to the case of no $U(1)$
gauge field first ($\a_0=0$ in Eq.\ \re{a1})).  As a consequence of the
reformulation we merely have to solve the massless Dirac equation with
the background metric $g$ (even despite the fact that $\gamma' \neq
0$). Here it is not so much decisive that $g$ is a space of constant
curvature, the main point is that we know its conformal factor {\em
explicitly}, cf.\ Eq.\ \re{const}). In such a case the solution of the
Dirac equation is the one of Minkowski space, i.e.\ again consisting
of right-- and left--movers, $\chi_R(x^+)$ and $\chi_L(x^-)$,
conformally  {\em transformed}\/ to the space with metric $g$ as a
field with conformal weight minus one half: \be \psi = \exp (-\xi/2)
\, \left(
\begin{array}{c} \chi_R(x^+) \\  \chi_L(x^-) 
\end{array} \) \pl{RLLsg} \,\, , \ee where, in the case under 
consideration, $\xi$ is given by Eq.\ \re{const}). (Note that the
result is no more a superposition of a left-- and a right--mover,
$\xi$ depending on $x^+$ {\em and}\/ $x^-$, but only a conformal
transform thereof).

Up to now the solution of the field equations was immediate. Now we
come to solving the equations \re{eomg2}), however, which is a less
trivial task (for $a \neq 0$). Due to the introduction of $g$ and
$\phi$ as new variables, there is no $V$-term and $U$ is linear. The
essential restriction of this section, Eq.\ \re{relation}), moreover,
ensures that the whole system becomes linear in $\phi$.  So we are
left with the following three equations \ba \nabla_\pm \nabla_\pm \phi
&=& {\bf T}_{\pm \pm} \pl{dil1} \\ e^{-2\xi} \6_+ \6_- \phi + 2a \phi
&=& -b \pl{dil2} \ea where $\xi_{,\pm} \equiv \6_\pm \xi$ and where we
used ${\bf T}_{+-}=0$ (as a consequence of $\CA=0$). The lefthand side
of the first two equations may be rewritten more explicitly as: $\(
\6_\pm - 2 \xi_{,\pm} \) \6_\pm \phi \equiv e^{2\xi} \6_\pm e^{-2\xi}
\6_\pm \phi$. (The only non--vanishing components of the Christoffel
connection are $\Gamma^+{}_{++}= 2\xi_{,+}$ and $\Gamma^-{}_{--}=
2\xi_{,-}$). The respective righthand sides of Eqs.\ \re{dil1}) are
given by \ba {\bf T}_{++} &=& {\bf T}_{++}(x^+) \equiv (f_+')^2 + i
\chi_R^\ast \dlr_+ \chi_R \, ,\label{T+} \\{\bf T}_{--} &=& {\bf
  T}_{--}(x^-) \equiv (f_-')^2 + i \chi_L^\ast \dlr_- \chi_L \,
.\label{T-} \ea Here the indices $+$ and $-$ are world sheet indices
($\m$, $\n$, $\ldots$), not to be mixed up with the frame bundle
indices ($a$, $b$, $\ldots$). 

Note that with both indices lowered (and only then!) the energy
momentum tensor ${\bf T}_{\m\n}$ is invariant with respect to the
conformal field redefinition \re{gneu}): ${\bf T}_{\m\n} =T_{\m\n}$.
Here the bold faced quantity is the energy momentum tensor of the {\em
  original}\/ theory \re{I}), given in Eqs.\ 
\re{Tscal},\ref{TU1},\ref{Tferm}), and $T_{\m\n}$ the energy momentum
tensor following from \re{J}) upon variation with respect to the
auxilary metric $g$.  This holds also for the $U(1)$ gauge field; the
fact that it is not conformally invariant in two dimensions, in
contrast to scalar fields, e.g., is reflected in that the
redefinition of $\a$ to $\wt \a$ contains the conformal factor, while 
the latter is absent in the analogous transition $\b\to \wt \b$ (cf.\ Eq.
\re{wtbeta})).

As ${\bf T}_{\pm \pm} = {\bf T}_{\pm \pm}(x^\pm)$ the solution of
\re{dil1},\ref{dil2}) is immediate for $a=0$, since then also $\xi
\equiv 0$ (cf.\ Eq.\ \re{const})); the complication arises when
allowing for $a \neq 0$ in Eq.\ \re{relation}).

The general local solution to the equations \re{dil1}, \ref{dil2}) is
of the form: \be \phi = \CT_+(x^+) + 2 \xi_{,+} \, \int^{x^+}\CT_+(z)
\, dz \, + \; (+ \lra -) \; - \, \frac{bx^+x^-}{1+a x^+x^-}
\pl{solution} \ee with \be \CT_\pm (u) = \int^udv\int^v {\bf T}_{\pm
  \pm}(z) \, dz + \2 K + k_\pm x^\pm \, , \pl{Tpm} \ee where in the
last line we displayed constants of integration $K$ and $k_\pm$
expli\-cit\-ly. There was no need to introduce two different constants
of integration instead of $K$ as within \re{solution}) they anyway
would contribute the same (i.e.\ there difference drops out and
without loss of generality the two constants may be set equal ---
assuming that the lower boundaries in the integrations are fixed in
some arbitrary way). The constants of integration $C_\pm$ from the
integral over $\CT_\pm$, on the other hand, may be absorbed into a
redefinition of $k_\pm$ for $a \neq 0$, $k_\pm \to k_\pm + C_\pm/a$,
while they disappear for $a=0$ due to $\xi_{,\pm} =0$; so we did not
display them.

That Eq.\ \re{solution}) is a solution of the coupled system
\re{dil1}, \ref{dil2}) may be established readily using two relations,
following from the definition of $\xi$ in Eq.\ \re{const}):
$\6_\pm \6_\pm \xi = (\xi_{,\pm})^2$ and $\6_+ \6_- \xi = - a \exp (2
\xi)$, the latter of which is equivalent to the statement that
$g=2\exp(2\xi) dx^+dx^-$ describes a space of constant curvature
$R(g)=-4a$.  These two equations are {\em essential}\/ for 
\re{solution}) to solve Eqs.\ \re{dil1}, \ref{dil2}). Thus despite the
simplicity and the apparent generality of our solution \re{solution}),
it solves the field equations only for the specific function $\xi$ 
defined above!  Having found a particular integral of the linear
equations \re{dil1}, \ref{dil2}), one is left with finding the general
solution of the homogenous system. As we will see right below, the
homogenous solutions $\phi_{hom}$ are indeed incorporated in
\re{solution}) taking into account the freedom in choosing the
constants $K$ and $k_\pm$.  This then concludes the proof that
\re{solution}) is the general solution of Eqs.\ \re{dil1},
\ref{dil2}).
 
To our mind the homogenous system is most transparent in a different
coordinate system, namely the one in which $g=2dx^0dx^1 + 2a \(x^0
dx^1\)^2$, resulting from \re{const}) by means of $x^-=x^1$ and $x^+ =
x^0/(1-ax^0x^1)$. These coordinates have 
 the advantage that the zero--zero component
of the homogenous system $\nabla_\m\6_\n \phi_{\hom} + g_{\m\n} \, (2a
\phi_{\hom} + b)=0$ reduces to $\6_0\6_0 \phi_{\hom} =0$, so that
$\phi_{\hom}$ is found to be at most linear in $x^0$! The $x^1$
dependence is then restricted by the remaining two equations, which
are $(\6_1 -2a x^0) \6_0 \phi_{\hom} +2a \phi_{\hom} =0$ and
$(1-x^0\6_0)\6_1\6_1 \phi_{\hom}=0$, where we have made use of the
former equation to simplify the one--one component of the field
equations to reduce to the latter equation. This then leads to \ba
\phi_{hom}&=& k_+ x^0 + k_- x^1 (1 - a x^0 x^1) + K (1 - 2a x^0 x^1)
\nonumber \\ &=& \frac{k_+ x^+ + k_- x^- + K (1 - a x^+ x^-)}{1 + a x^+x^-}
\pl{hom} \ea for three free constants $k_\pm$ and $K$. As the
notation already suggests, they indeed coincide with the three
constants found in \re{solution}).

As remarked already in Sec.\ 1, in the absence of matter fields and
with a spacetime--topology $\dR^2$, the general solution of the field
equations modulo gauge symmetries is parametrized by {\em one}\/ real
quantity only (cf.\ \cite{Classgendil}). As the above three constants
$K$ and $k_\pm$ remain in the matterless (``vacuum'') solution, not
all of them can describe physically (or geometrically) different
spacetimes.  Indeed, Eq.\ \re{const}) has a residual gauge freedom:
\be x^\pm \ra \l^{\pm 1} \, \frac{x^\pm + s^\pm}{1 - s^\mp a (x^\pm +
  s^\pm)} \, , \pl{resi} \ee parametrized by three real constants $\l
\neq 0$ and $s^\pm$. For $s^\pm :=0$, e.g., this is nothing but the
rescaling $x^+ \ra \l x^+$, $x^- \ra x^-/\l$, leaving \re{const})
invariant at first sight; by means of Eq.\ \re{hom}) it leads to the
identification $(k_+,k_-) \sim ((k_+/\l), \l k_- )$. A more detailed
analysis shows that the factor space of $K$ and $k_\pm$ modulo the
action induced on them by Eq.\ \re{resi}) is indeed a one parameter
space. (E.g., for $a=0$, $b \neq 0$, and no matter fields,
$T_{\m\n}=0$, one easily establishes that $k_\pm$ may be put to zero
by shifts in $x^\pm \to x^\pm + s^\pm$; thereafter rescalings with
$\l$ have no effect anymore, so that $K$ remains as a physical
parameter. --- If, on the other hand $a=b=0$ and there are no matter
fields, then one can achieve, e.g., $k_+ = 1$ and $K=0$, with $k_-$
remaining.) In many cases the remaining parameter may be given the
interpretation of the ``mass'' $M$ of the spacetime described by
$\gtrue$. Strictly speaking, the above consideration applies to the
matterless case (reconsidered in the present gauge). However, we think
that the present {\em counting}\/ of gauge--invariant parameters will
be unmodified when adding matter, so that also in the general case only
one of the three parameters $k_\pm$ and $K$ will survive.

In summary we get the following results: Combining the solution
\re{solution}) with Eqs.\ \re{gneu}), (\ref{const}), and
(\ref{relation}), the metric is brought into the form of Eq.\ 
\re{hihi}), while $\Phi=U^{-1}(\phi)$. The scalar field is given by
Eq.\ \re{Lsgphi}), the fermionic field by solving Eqs.\ 
\re{psi},\ref{RLLsg}) for $\bfPs$. The local solutions are
parametrized by the choice of the one--argument functions $f_\pm$ and
$\chi_\pm$ as well as by the one constant of the
vacuum theory (cf.\ the discussion above). In terms of these
data the energy momentum tensor, as defined in Eqs.\ 
\re{Tscal},\ref{Tferm}), has the form \re{T+},\ref{T-}) (while ${\bf
  T}_{+-} =0$).

\vskip5mm

{\centerline{\em Inclusion of gauge fields (arbitrary $\a_0$):}}

\vskip2mm

We now turn to the system with an additional $U(1)$ gauge symmetry.
Assuming $\a$ in the action \re{U1}) to be subject to the constraint
\re{a1}), we obtain a constant coupling $\wt \a =\a_0$ in the
reformulated action \re{J}). The matter sector of the theory still
poses no problem, as it is mapped to a corresponding system on
a space of constant curvature. So we are left with the dilatonic
equations \re{eomg2}), or, equivalently, with: \ba e^{2\xi} \6_\pm
e^{-2\xi} \6_\pm \phi &=& {\bf T}_{\pm \pm} \pl{dil1} \\ \6_+ \6_-
\phi - 2 (\6_+ \6_- \xi) \phi &=& -b e^{2\xi} - {\bf T}_{+-} \, .
\pl{dil2neu} \ea For ${\bf T}_{+-}=0$ this  takes the form of
Eqs.\ \re{dil1},\ref{dil2}). However, the present ${\bf T}_{\m\n}$ has
decisive differences to the previous considerations: First, ${\bf
  T}_{++}$ and ${\bf T}_{--}$ depend on {\em both}\/ coordinates $x^+$
and $x^-$ now (otherwise we would be forced to $\CF=0$ and,
consequently, also to $\bfPs=0$, cf.\ the remarks following Eq.\ 
\re{eomA})!). Second, also ${\bf T}_{+-}=-(\a_0/2) \, \exp(-2\xi) \, \[
\CF_{+-}\]^2 \neq 0$ for $\CF \neq 0$; i.e.\ due to the presence of
the $U(1)$ field the energy momentum tensor is no more tracefree 
except for $\bfPs = 0 = \CF$. 

The form of ${\bf T}_{\m\n}$ is still restricted by some
decisive relations. These may be expressed most elegantly when using
the trace $T$ of the energy momentum tensor with respect to the
auxiliary metric, $T =g^{\m\n}T_{\m\n}=g^{\m\n}{\bf T}_{\m\n}$. (We
remind the reader that in view of Eq.\ \re{gneu}) ${\bf T} = \O^2 \, T
\neq T$ in general). The energy momentum tensor satisfies:
\ba \6_+ \6_- \sqrt{|T|} &=&0
\label{xv1} \\ \6_\pm T&=& - 2\exp(-2\xi) \, \6_\mp {\bf T}_{\pm\pm} 
\label{xv2} \ea The first of these relations follows from Eq.\ \re{eomA}). 
It allows to write $T=T^{U(1)}=-\a_0 \, \[ \exp(-2\xi) \,
\CF_{+-}\]^2$ in the form $-[F_+(x^+)+F_-(x^-)]^2$, where, again
through Eq.\ \re{eomA}), the functions $F_\pm$ are determined by the
fermion fields up to an additive constant.  The second relation, Eq.\ 
\re{xv2}), is equivalent to $\nabla^\m {\bf T}_{\m\n} =0$, where this
equation is understood entirely with respect to the  {\em auxiliary}\/
metric $g$; it follows from the equations fulfilled by
the matter fields, and, simultanously, it is the integrability condition
of the dilatonic system of differential equations. Note that the
second relation allows to express ${\bf T}_{++}$ and ${\bf T}_{--}$ in
terms of $T$ and $\xi$ up to a function of the single
variable $x^+$ and $x^-$, respectively: 
\be {\bf T}_{++}=-\2 \int_0^{x^-} \, 
e^{\left(2\xi( {x}^+,{\wt x}^-) \right)} \, 
(\6_+ T)( {x}^+, {\wt x}^-) \, d {\wt x}^- 
+ {\bf T}_{++}(x^+,0) \, \ee
with a similar equation for ${\bf T}_{--}$. The functions 
${\bf T}_{++}(x^+,0)$ and ${\bf T}_{--}(0,x^-)$ 
may then be eliminated from the right--hand side of the 
dilatonic system of linear differential equations by a particular 
integral of the type \re{solution}) found already above. One then is
left with finding a particular integral of the remaining system, with
inhomogeneities determined solely by  $T=-[F_+(x^+)+F_-(x^-)]^2$. 

This still turns out to be quite a hard problem. After the dust
clears, however, it is possible to put the solution to the system
\re{dil2neu}) into the following form: \ba \phi&=& \int_0^{x^+}\! d\wt
x^+\int_0^{x^-} \! d\wt x^- \; \left(\frac{a(x^- -\wt x^-)(1+a\wt x^+
    x^-)}{1+a\wt x^+\wt x^-} -1\right)\; {\bf T}_{+-}(\wt x^+,\wt x^-)
\, - \nn && -\, \frac{bx^+x^-}{1+a x^+x^-}\, + \, \int^{x^+} \! d\wt
x^+ \int^{\wt x^+} \! du \; \frac{\(1+a\,ux^-\)^2}{\(1+a\, \wt x^+
  x^-\)^2} \; {\bf T}_{++}(u,0)\, +\,\nn &&+\, \int^{x^-} \! d\wt x^-
\int^{\wt x^-} \! dv \; \frac{\(1+a\,x^+ v\)^2}{\(1+a\, x^+ \wt
  x^-\)^2} \; {\bf T}_{--}(0,v) \,+\, \phi_{hom}\; . \label{phiU1} \ea
Here $\phi_{hom}$ is the homogenous solution \re{hom}), containing the
one gauge--invariant parameter of the vacuum theory that is left over when
taking into account the residual gauge freedom \re{resi}) discussed
above. 

It is a somewhat cumbersome calculation to show that \re{phiU1})
indeed solves the system \re{dil2neu}) and we will not provide any
further intermediary steps here so as to not become too technical.
Actually, it is even not completely obvious to see how the above $\phi$
reduces to the previous form \re{solution}) for ${\bf T}_{+-}=0$ and
${\bf T}_{\pm\pm}={\bf T}_{\pm\pm}(x^\pm)$. We recommend this
consistency check as an exercise to the reader.

Up to gauge transformations, the general local solutions are
parametrized by the choice of the initial data for the matter fields
$f$ and $\bfPs$, as well as by the gauge invariant constant contained
in $\phi_{hom}$ as well as one further constant of integration in the
field strength $\CF$.

\vskip7mm

In this section we analysed 2d gravity--matter models in which the
{\em auxiliary}\/ metric $g$, defined in Eq.\ \re{gneu}), has three
Killing vectors (cf.\ Eq.\ \re{const})). Basically this was the
defining restriction for the class of models considered in this
section. We remark that for a completely general model \re{I}) it is
also possible to find {\em those}\/ solutions, in which $g$ has one
Killing field (stationary/homogenous auxiliary spacetimes); for this
special {\em subclass}\/ of solutions of the general model, one can
reduce the field equations to one ordinary differential equation of
third order. We refer to \cite{Diplom} for details.

\section{Chiral Solutions}
In this section we want to show that chiral fermions coupled to 
a general gravity action \re{gdil}) and to gauge fields 
is a classically solvable system. To start with, we will, however, 
discuss the fermion--gravity system with both chiralities,
disregarding, furthermore, possible gauge fields in a first step; 
thus to begin with, our Lagrangian has the form  
$I=I_{gdil} + I_{ferm}$.

\vskip10mm

{\centerline{\em Chiral fermions coupled to generalized dilaton gravity:}}

\vskip2mm

As remarked already in the introductory section, $I_{gdil}$ may be
formulated equivalently by means of the action \re{PS}), with
$I_{ferm}$ being given by the expression \re{AJ}). This comes about as
follows (cf.\ also \cite{PartI}): As shown in the previous section,
appropriate field redefinitions brought $I$ into the form of Eq.\ 
\re{J}) with $\wt \a = 0 = \wt \b$. Now we want to reexpress this
action for $g$, $\phi$, and $\psi$ in its Cartan formulation, using an
(auxiliary) zweibein $e^a$ with $g=2e^+e^-$ and an (auxiliary) spin
connection $\omega$. In this context it is important to note, however,
that the zero torsion condition for $\o$ does not result automatically
from the variation of the 2d gravity action as it would in the 4d
Einstein Hilbert action.  Consequently, using $\omega$ as an
independent variable, we need Lagrange multiplier fields, $X^\pm$, to
enforce zero torsion $De^a\equiv de^a + \e^a{}_b \, \o \wedge e^b =0$.
The resulting gravitational part of the action,\footnote{Here we use
  conventions $\e^{+-}=1$ and $\sqrt{-\det g}\,d^2 x = e^+ \wedge
  e^-$.}  $\int_{\CM} \phi d \omega + X_{a}De^{a} + \wt W(\phi)e^- \wedge
e^+/2$, may be rewritten identically as given in Eq.\ \re{PS}), if we
collect fields according to $A_i \equiv (e^+,e^-,\o)$ and $X^i \equiv
(X^-,X^+,X^3)$ and define the matrix ${\cal P}^{ij}$, $i,j \in
\{-,+,3\}$, by \be ({\cal P}^{ij})=\(
\begin{array}{ccc}
  0 & -\2 \wt W(X^3) & -X^{-}\\
  \2 \wt W(X^3) & 0 & X^{+}\\
  X^{-} & -X^{+} & 0\\
\end{array} \). \pl{P}
\ee After application of the 2d identity $\det (e_\r^b) \, e_{a}^{\m}=
\e(ab)\e(\m\n) \, e_\n^b$, where $\e(\cdot \, \cdot)$ denotes the 2d
antisymmetric symbol (without any metric dependence), the fermion part
of the action, on the other hand, takes the form \re{AJ}) with $J^j_\m
= (B^-_\m,-B^+_\m,0) = (i\psi_R^\ast {\stackrel{\lra}{\6}}_\m \psi_R,
-i\psi_L^\ast {\stackrel{\lra}{\6}}_\m \psi_L,0)$, where $\psi_{R,L}$
denote the posi\-tive/negative chirality components of $\psi$. Thus, up
to an irrelevant factor, the total action $I=I_{gdil} + I_{ferm}$
becomes
\be I= \int_{\CM} A_i \w (dX^{i}+J^{i})+\2 \CP^{ij}A_i\w A_j
\pl{PSJ} \ee with $i,j \, \in \, \{-,+,3\}$ and $J$ as
given above. As we stay on the purely classical level within this
paper, the fermionic variables may be taken commuting and $J$ may be
simplified by the following parametrization: \be \psi \equiv \(
\begin{array}{c} \psi_{R}\\ \psi_{L}
\end{array} \) := 
\( \begin{array}{c}
r \, \exp(-i \rho)\\l \, \exp(i \lambda)
\end{array} \)  \quad \Ra \quad  
J^i = (r^2d\rho, l^2 d\lambda, 0) \; . \pl{Jpar}
\ee

Following the remarks in the opening section, a change of
variables in the target space of the theory, $X^i \ra \wt X^i$,
inducing a change of variables $A_i \ra \wt A_i = (\6  X^k/\6 \wt X^i)
\, A_k$, may be used to simplify the tensor ${\cal P}^{ij}$. In
particular, as a consequence of the Jacobi identity \be {\cal P}^{il}
\, \6_l \CP^{jk} + \mbox{cyclic}(ijk) =0 \pl{Jakobi} \ee fulfilled by
${\cal P}^{ij}$ as given in Eq.\ \re{P}) above, there always exist
coordinates $\wt X^i$ such that $\wt {\cal P}^{ij}$ takes (constant)
Casimir Darboux (CD) form and the action becomes \be I=\int_{\CM} \wt
A_i \w (d \wt X^{i}+\wt J^{i})+\wt A_2\w \wt A_3 \plabel{ICD} \ee 
with $i \in \{ 1,2,3\}$.

Note, however, that now the origianally simple currents \re{Jpar}) 
may have become complicated in the transition   $J^i
\ra \wt J^i \equiv (\6 \wt{X^i}/\6 X^a)\, J^a$. 

A possible choice of CD coordinates of \re{P}) is provided by
\cite{PartI} \be \wt X^i = (2X^+X^--\int^{X^3}\wt W(t)dt, \ln |X^+| ,
X^3) \pl{CD} \q .  \ee This induces the follwing form for the current:
\be \wt J^i = (2X^+J^-+ 2X^-J^+\, , \; J^+/X^+,0) \,\, , \pl{wJ} \ee
where $X^\pm$ are to be understood as functions of the new variables
$\wt X^i$.  While in the case of pure gravity, $J^i \equiv 0$, also
the last of the initially three potentials, $\wt W(\phi)$, could be
eliminated from the action by means of the above change of variables,
it now creeps in again through $\wt J^i$. {\em However}, from Eq.\ 
\re{wJ}) it is obvious that $\wt W(X^3)$
can be eliminated in the case of {\em chiral}\/ fermions:
$\psi_L=0$ implies $J^+=0$ and $\wt W$ drops out of \re{ICD})! With
the further (Lorentz invariant) field redefinition \be \wt r := 2X^+
\, r^2 \pl{wr} \ee the transform of the chiral current $J^i =
(J^-,0,0)$ is $\wt J^i = (\wt r d\rho,0,0)$ and the action \re{ICD})
takes the trivialized form \be I_{chiral}=\int_{\CM} \ti A_{i}\w d\ti
X^{i}+\ti r\ti A_1\w d\r+ \ti A_2\w \ti A_3 \pl{DA.3.81} \; .  \ee We
are thus left with solving the field equations of this very simple
action.

Variation with respect to $\wt A_2$, $\wt A_3$ yields \be \wt A_2=d\wt
X^3\; ,\quad \wt A_3=-d\wt X^2\; , \pl{DA.3.82} \ee which may be used
to eliminate these fields together with $\wt X^2$, $\wt X^3$. This
simplifies \re{DA.3.81}) further to \be I_{chiral}=\int_{\CM} \ti A_1\w
\(d\ti X^1 + \ti r d\r \)\; .  \pl{DA.3.84} \ee According to
\re{DA.3.84}), $\wt A_1$ is exact. Thus, with an appropriate choice of
coordinates, $\wt A_1 = dx^1/2$. This may be used to establish that all
the remaining fields $\rho$, $\wt r$, and $\wt X^1$ are functions of
$x^1$ only and the latter is determined up to a constant of
integration $c$ by the former two via \be \wt X^1= \wt X^1(x^1)=-\int \wt
r(x^1)\rho'(x^1) dx^1 + c \,\, \pl{solX1} . \ee The gauge freedom of our
gravity theory may be fixed completely by choosing $\wt X^3$ as the
second coordinate $x^0$ and by using a Lorentz frame such that $X^+ =1
\Rightarrow \wt X^2=0$. 

Thus in the present framework the field equations
could be trivialized. The local solutions are parametrized by the
choice of initial data for $\wt r(x^1)$, $\rho(x^1)$, and the
integration constant $c$, furthermore. The latter is the only
parameter remaining in the matterless case, in coincidence with
the literature \cite{Classgendil}.

We finally reexpress the solution in terms of our original variables
$\gtrue$, $\Phi$, and $\bfPs$, i.e.\ we perform the transformation
inverse to the one that has led us from the two parts \re{gdil}) and
\re{ferm1}) of the original action $I$ (with the additional constraint
$\bfPs_L = 0 \LRA \psi_L =0$) to the action \re{DA.3.81}) or
\re{DA.3.84}). In the gauge chosen above the dilaton has the form
$\Phi =U^{-1}(x^0)$. The chiral fermion field is given by (cf.\ Eq.\ 
\re{psi})) \be \bfPs_{R}= \sqrt{\frac{\wt
    \O(x^0)}{|\gamma(U^{-1}(x^0))|}} \psi_R(x^1) \;\, , \qquad
\psi_R(x^1) \equiv r(x^1)\exp{\left[-i\r(x^1)\right]} \,\; ,\pl{rf}
\ee where $\wt \O \equiv \O \circ U^{-1}$. By means of \ba A_-&=&2\wt
A_1 = dx^1 \, , \nn A_+ &=& \exp(-\wt X^2) \left[ \wt A_2+ \(\wt X^1 +
  \int^{\wt X^3} \wt W(t) dt \right) \, \wt A_1 \right] \, = \nn {}&=&
dx^0 + \2 \(\wt X^1 + \int^{x^0} \wt W(t) dt \right) \, dx^1 \, ,
\pl{Umrechnung} \ea furthermore, the physical metric $\gtrue = g/\wt
\O^2 (x^0)= 2A_+A_-/\wt \O^2 (x^0)$ (cf.\ Eq.\ \re{gneu})) becomes \be
\gtrue=\frac{1}{\wt \O^2(x^0)}\left[2dx^0 dx^1+\left(-2\int^{x^1} {\bf
      T}_{11}(u)du +c+\int^{x^0}\wt W(t)dt\right)(dx^1)^2\right] \, .
\pl{gwahr} \ee Here we rewrote Eq.\ \re{solX1}) by means of ${\bf
  T}_{11}=r^2(x^1)\rho'(x^1)$, which follows directly from Eqs.\ 
\re{Tferm},\ref{gneu},\ref{psi}).  We remark that ${\bf T}_{11}$ is
the {\em only}\/ nonvanishing component of ${\bf T}_{\m\n}$ here, and
that it has support along null lines ($x^1=const$ is null according to
Eq.\ \re{gwahr})).

A further coordinate change $x^0 \ra \int^{x^0} dt/\wt \O^2(t)$ brings
the metric \re{gwahr}) into the form \re{CEF}), announced in Sec.\ 1;
the functions $h$, $k_0$, and $h_1$ may be easily identified from the
above.

\vskip5mm 

{\centerline{\em Chiral fermions coupled to gravitational actions
  with non--zero torsion:}}

\vskip2mm 

We briefly discuss changes that occur, if one allows for gravitational
actions with torsion terms. Such actions result \cite{PartI}, if one
allows the potential $\wt W$ to depend also on $X^+X^-$ in addition to
$X^3$ in Eq.\ \re{P}). E.g., the Katanaev--Volovich model \cite{KV}
results from $\wt W(2X^+X^-,X^3)=-\a X^+X^- -(X^3)^2+\L/{\a }^2$ for
two real constants $\a$ and $\L$.  In this case $g$ is already the
true (physical) metric, without any additional conformal
transformation. Now CD coordinates have the form \be \wt X^i = (\,
C(2X^+X^-,X^3) \, , \, \ln |X^+| \, ,\, X^3) \, ,\pl{CD2} \ee where
the two--argument function $C(u,v)$ is a solution to the differential
equation $\wt W(u,v) \6_u C + \6_v C=0$. For at most
linear $(X^+X^-)$--dependence of $\wt W$ (such as is the case for the
KV--model), $C$ may be determined explicitly from this differential
equation (cf.\ \cite{PartI}). However, to determine the general
solution of the field equations, this is not necessary; it may be
written in terms of the function $C$. Actually, using \re{CD2}), {\em
  all}\/ the steps and formulas from \re{ICD}) to \re{solX1}) apply
also in this case, except for \re{wJ}), which generalizes to $\wt
J^1=2(\6_u C) \, \left[\exp (\wt X^2)J^-+X^-(\wt X)J^+\right]/2$,
where the second term drops out upon restricton to chiral solutions
$J^+ \equiv 0$.  The Casimir function $C$ enters only when determining
the metric from the fields $\wt A_i$. E.g., $g$ now becomes \be
g=2\6_uC \, \left[dx^0dx^1+ X^- \, \6_uC \, (dx^1)^2 \right]\;\, , \ee
where $\6_uC$ and $X^-$ are to be understood as functions of the
CD--coordinates \re{CD2}), which, as before, are given by Eq.\ 
\re{solX1}) and $\wt X^2 = 0$, $\wt X^3=x^0$.

In the particular case of the KV--model the above formulas reproduce
the results found in \cite{Kummer,Solodukhin}. Note that if instead of
$J^+ \equiv 0$ we put $J^- \equiv 0$, one still may proceed as above,
merely replacing $\wt X^2 = \ln|X^+|$ by $\wt X^2 = -\ln|X^-|$.  

In \cite{Solodukhin} it has been claimed that the general solution for
fermions of {\em both}\/ chiralities may be obtained ``in the same
manner'' as those of one chirality. We did not succeed to verify this
(cf.\ also the remarks around Eq.\ \re{wJ})).  It would be very
interesting to see this general solution of the KV-model coupled to
fermions of both chiralities. In particular, from the present
perspective it seems most likely that such a solvability would
generalize to the {\em whole class}\/ of models \re{gdil}) with the
same matter content.

 
\vskip5mm 

{\centerline{\em Chiral fermions coupled to generalized dilaton gravity
  and a $U(1)$ gauge field:}}

\vskip2mm 

We now turn to the discussion of additional gauge fields. We start
with $U(1)$ gauge fields, taking the gravitational part torsion free
for simplicity, $I=I_{gdil} + I_{ferm} + I_{U(1)}$, Eqs.\ 
\re{gdil},\ref{ferm1},\ref{U1}). In \cite{PartI} it has been shown
that a gravity Yang--Mills system of the above kind (but without the
matter contribution $I_{ferm}$) may be described by a Poisson
$\s$--model \re{PS}) where the indices $i$ run from one to $d+3$, $d$
being the dimension of the structure group $G$ of the Yang--Mills
theory. In the present case of the abelian group $G=U(1)$ we have
$d=1$ and the Poisson tensor ${\cal P}^{ij}$ has the same form as
\re{P}) with the addition of a forth row and line with zeros and the
replacement $\wt W(X^3)$ by $\widehat{W}(X^3,X^4)= \wt W(X^3)-(X^4)^2/
\wt \a (X^3)$. This form of the total gravity--$U(1)$--action comes
about when bringing the $U(1)$--action, Eq.\ \re{U1}) or better the
last term of Eq.\ \re{J}), into first order form: \be I_{U(1)}= \int
{\CA} \wedge dE + \frac{1}{2 \wt \alpha} E^2 e^+ \wedge e^- \, ; \ee
this yields the previous form of the $U(1)$--action upon elimination of the
``electric field'' $E$ by means of its equations of motion. (The
latter is $E=-\wt \a \ast  dA$, if $\ast$ denotes the Hodge dual
with respect to $g$, or, expressed in the original variables, $E= - \a
\ast dA$, where $\ast$ denotes the Hodge dual with respect to
$\gtrue$, $\a$ and $\wt \a$ being related through Eq.\ \re{wtbeta})).
The gravity--$U(1)$--action then takes Poisson $\s$--form with the
Poisson tensor as described above and the identifications $A_4 = \CA$ and
$X^4=E$ in addition to those for $A_i$ and $X^i$, $i \in \{-,+,3\}$, 
made  in the absence of the gauge field. Alternatively,
$A_4$ may be defined also as in  Eq.\ \re{A4}) below. This will turn
out to be more convenient in the case of chiral fermions, while it
does not change the form of $\CP^{ij}$. For the moment, however, we
will stick to $A_4=\CA$, as in the matterless case discussed in
\cite{PartI}.

As we saw already in Sec.\ 2, in contrast to the spin connection, the
$U(1)$ connection $\CA$ does not drop out from $I_{ferm}$ (cf., e.g.,
Eq.\ \re{Tferm})).  The additional contribution in $I_{ferm}$
containing the connection one--form is $\int \, \g (\Phi) \,
(-\bfPs_R^*\bfPs_R {\bf e}^+ \newline + \bfPs_L^*\bfPs_L {\bf e}^-) \wedge
{\CA} = \int (-\psi_R^*\psi_R A_- +\psi_L^*\psi_L A_+) \wedge A_4$
(using $A_4=\CA$).  Although this is of the form $A_i \wedge A_j$ with
an appropriate coefficient matrix, it is {\em not}\/ advisable to
incorporate it in ${\cal P}^{ij}$. The reason is that the thus
redefined tensor $\cal P$ would no more satisfy the Jacobi identity
\re{Jakobi})!  Since the latter is at the heart of our approach, we
proceed differently: With the currents \be J^i = (r^2 \, (d \rho -
{\CA}), l^2 \, (d\lambda + {\CA}), 0,0) \pl{J4}\, , \ee where we used
the parametrization \re{Jpar}) again, the coupled system takes
the form \re{PSJ}), with $i,j$ running over four values now.

The current \re{J4}) suggests to change variables from
$\CA$ to the gauge invariant combination $\CA - d \rho$. We thus
(re)define $A_4$ by: \be A_4 := {\CA} - d \rho . \pl{A4} \ee In the
case of one chirality, $\bfPs_L := 0 \LRA l=0$, to which we will
restrict ourselves in the following, the only non--vanishing component
of the current becomes $J^{-} = -r^2 A_4$ then, while the action again
takes the form \re{PSJ}) (with the same four times four matrix $\CP$
as above).  The field $\rho$ is now seen to drop out completely from
the action as a total divergence (exact two--form).  This is a
manifestation of the $U(1)$ gauge invariance. Dropping the total
divergence and implicitly the phase $\rho$ of the fermion field,
eliminates the gauge freedom and, at the same time, saves us the study
of the associated redundant field equations (according to Noether's
second theorem any local symmetry gives rise to a relation among the
equations of motion, cf., e.g., \cite{Sundermayer}).

Next we transform the action to CD coordinates.  The extended Poisson
tensor has CD coordinates of the form \re{CD}) with two changes:
First, $\wt W(X^3)$ is replaced by $\widehat{W}(X^3,X^4) \equiv \wt
W(X^3)-(X^4)^2/ \wt \a (X^3)$, $X^3$ being the variable to integrate
over, and second there is a fourth coordinate, $\wt X^4:= X^4$, which
now is the second Casimir of the (four by four) matrix $\CP$ beside
$\wt X^1$. Introducing again the Lorentz invariant function \re{wr})
and using Eq.\ \re{DA.3.82}) to get rid of some irrelevant field
equations\footnote{Here again one drops a total divergence, getting
  rid of the local Lorentz symmetry and half of the diffeomorphism
  invariance. Certainly the resulting action, Eq.\ \re{neuW}) below or
  Eq.\ \re{DA.3.84}) above, respectively, are still invariant under
  the full diffeomorphism group (as is obvious from its formulation in
  terms of forms without using any background metric); the apparent
  paradox is resolved by noting that the solution for the remaining
  fields depend on one coordinate function only ($x^1$ with our choice
  of coordinates), so that the part of the diffeomorphism group
  eliminated by means of \re{DA.3.82}), which is $x^0 \ra x^0(\bar
  x^\mu)$, acts trivially on these solutions.}, we end up with \ba \wt
I=\int_{\CM} \wt A_1\w d\wt X^1+\wt A_4\w \(d\wt X^4+\wt r \wt A_1\)
\pl{neuW} \q , \ea where $\wt r$ is defined as without gauge field.
The field equations of the above action are found and solved readily.
Thereafter one transforms back to the original variables.  We
display the result of these considerations only. In an appropriate
gauge one finds $\Phi=U^{-1}(x^0)$, $r=r(x^1)$, and $\rho=\rho(x^1)$,
as before.  $\CA$ has the form: \be \CA=-E(x^1) \( \int^{x^0}
\frac{dz}{\wt \a(z)} \) \, dx^1 \, , \pl{ALsg} \ee where $E=E(x^1) = -
\int^{x^1} r^2(z) dz+ \widehat{c}$. The metric, finally, is given by
\ba \gtrue&=&\frac{1}{\wt \O^2(x^0)}\left[2dx^0 dx^1+ \left(
    \int^{x^0}\wt W(z)dz - 2 \int^{x^1} r^2(t)\r^{\prime}(t)dt + c -
  \right.\right.\nn &&\qq \left.\left.- \, E^2(x^1)\, \int^{x^0}
    \frac{dt}{\wt \a(t)} \, \right) (dx^1)^2 \right]\pl{gu1} \q .  \ea
We recall that $\wt \O \equiv \O \circ U^{-1}$ and $\wt \a = \wt \O^2
\cdot (\a \circ U^{-1})$, where the function $\O$ is given by Eq.\ 
\re{gneu}). 

The generalization to non--trivial torsion is straightforward.


\vskip8mm 

{\centerline{\em Multiplet of chiral fermions with YM-fields and 
generalized dilaton gravity:}}

\vskip2mm 

The above results may be generalized also to gauge fields of an
arbitrary non--ablian structure group $G$ with a chiral fermion
multiplet $\bfPs_R$ in the fundamental representation of $G$.  $G$ is
assumed to allow for a non--degenerate ad--invariant inner product on
its Lie algebra, which we will denote by $tr$.  The kinetic term for
the Lie algebra valued gauge field $\CA$ is of the form $I_{YM} =
-\int_{\CM}\a(\Phi) \, tr(\CF\w *\CF)/2$, where $\CF = d\CA +{\CA}\w
{\CA}$ is the standard field strength and again we may allow for a
dilaton dependent coupling $\a$. As shown in detail in \cite{PartI},
$I_{YM}+I_{gdil}$ may be brought into the form \re{PS}), with a
(dim$G+3$)--dimensional target space.  Again all of the matter part of
the action may be formulated as $\int A_i \wedge J^i=\int A_- \w J^-$,
where now \be J^-=\gamma(\Phi) \, \left(\bfPs_R^\dagger \,
  (i\stackrel{\lra}{d}-\CA) \bfPs_R\right)\ee with $2\!
\stackrel{\lra}{d} \, \equiv \,
\stackrel{\ra}{d}\!-\!\stackrel{\la}{d}$.  So the total action
$I_{gdil}+I_{YM}+I_{ferm}$ is again of the form \re{PSJ}) (even in the
non--chiral case, but then also with a non--vanishing component $J^+$
containing the left--handed fermion multiplet).

As CD--coordinates (on the target space) we may choose: $\wt X^1 =
2X^+X^--\int^{X^3}\wt W(t)dt + tr(E^2) \, \int^{X^3} (1/\wt
  \a(t)) \, dt$, where $\wt \a$ is related to $\a$ via Eq.\
  \re{wtbeta}), $\wt X^2=ln|X^+|$, $\wt X^3=X^3$, while the following
CD--coordinates depend on the (Lie algebra valued) electric fields $E$
only --- we do not need to choose the latter functions explicitly to
solve the coupled system. In the fermion--YM--sector of the theory we
found it most advisable to switch between CD--adapted fields and the
original matrix valued fields $\CA$ and $E$. 

We do not want to go into the calculational details here. Still we
warn of a possible pitfall in using the present formalism: In view of
\re{PSJ}) when written with CD--adapted fields, one is tempted to
conclude $d\wt X^1 + \wt J^1=0$; this equation is {\em wrong},
however!  The reason for this failure is the following.  Expressing
$\wt J^1=2X^+J^-$ in terms of CD--adapted fields, from $\CA$ one picks
up a contribution proportional to 
$\wt A_1$: $\wt J^1 = \widehat J + 4 X^+ \gamma (\bfPs_R^\dagger E
\bfPs_R) \, \int^{X^3} \frac{dt}{\wt \a(t)} \, \wt A_1$.  This
contribution cancels from the action: $\int \wt A_1 \wedge (d\wt
X^1+\wt J^1) + \ldots =\int \wt A_1 \wedge (d\wt X^1+\widehat J) +
\ldots$. The correct field equation stemming from the variation with
respect to $\wt A_1$ is $d\wt X^1 + \widehat J=0$.  As may be seen,
$d\wt X^1 + \wt J^1=0$ would imply $\CF=0$; this is already wrong in
the abelian case, discussed in detail before.

At the end of the day one finds the following solution as a
straightforward generalization of the abelian results: The metric
$\gtrue$ may be brought into the form of Eq.\ \re{gu1}), where $E^2$
generalizes to $tr E^2$ and $r^2 \rho'$ to $i\psi_R^\dagger \dlr
\psi_R$. Here $\psi_R$ and $\bfPs_R$ are related by Eq.\ \re{psi}),
where again $\psi_R=\psi_R(x^1)$, in an appropriately chosen gauge,
and $\Phi= U^{-1}(x^0)$. $E$ is given by $E=E(x^1)=-\int^{x^1}
(\psi_R^\dagger(u) \, T^I \,\psi_R(u)) du \, T^I + \widehat c$, where
$T^I$, $I=1,\ldots, \mbox{dim}G$, denote the generators of the Lie
algebra and $\widehat c$ is some constant of integration restricted to
the Cartan subalgebra. $\CA$ takes the form of Eq.\ \re{ALsg}),
reinterpreted as a Lie algebra valued equation. The local
solutions are parametrized by the functions $\psi_R(x^1)$ and the 
$r+1$ ``vacuum parameters'' $c$ and $\widehat c$, $r$ being the rank
of the Lie algebra.

\section{Self--dual Scalar Fields}
In this section we study the (anti--)self--dual sector of the system
Eqs.\ \re{gdil}, \ref{scal}), or, equivalently, of the first line of
Eq.\ \re{J}). While the first two terms of the latter equation may be
described by \re{PS}), in the following we will bring also the action
of the scalar field into a compatible first order form. This will
finally allow us to map the (anti--)self--dual sector of the present
system to the chiral Lagrangian of the previous section. In this
process we will pick up an additional constraint, however. The latter
will enforce $\b=const \,$ so as to allow for non--trivial solutions.
With an appropriate conversion of symbols, these non--trivial
solutions may then be {\em copied}\/ from the solutions found in the
previous section, without the need of solving any field equation.  At
the end it will turn out among others that the metric again may be
brought into the form \re{gwahr}) where now ${\bf T}_{11}$ denotes the
only non--vanishing component of the energy momentum tensor of the
self--dual scalar fields.

We first bring the action for a scalar massless field into first order
form\footnote{The minus sign in front of the first integral is a
  consequence of $\e = - \sqrt{-\det g}\, d^2x$, following from the
  conventions chosen in this paper (first footnote of Sec.\ 4).}: \be
I_{scal}=-\2 \int_{\CM} \b(\Phi) df \wedge \ast df \cong \int_{\CM} B
\wedge df - \frac{1}{2\b(\Phi)} B \wedge \ast B \pl{IB} \, \, , \ee
where we introduced a one--form $B$ that equals $\beta(\Phi) * d f$ on
shell.  `$\ast$' denotes the Hodge dual operation with respect to the
dynamical metric $\gtrue$. However, due to the conformal invariance of
the action, and the fact that $f$ carries conformal weight zero, we
may equally well take $g$ instead, defined in Eq.\ \re{gneu}). Next we
split $B$ into its self--dual and its anti self--dual part. Due to
$\ast A_\mp = \pm A_\mp$, which is equivalent to $\ast {\bf e}^\pm =
\pm \ast {\bf e}^\pm$, this splitting is achieved by means of the
decomposition: \be B = R A_- + L A_+ \pl{B} \,\, , \ee where the
notation $R$ and $L$ stands for ``right--moving'' and
``left--moving'', respectively; in particular, the above $R$ has
nothing to do with the curvature scalar of $g$ or $\gtrue$. Combining
Eqs.\ \re{B}) and \re{IB}) and using $\b(\Phi)=\wt \b(\phi)$, Eq.\ 
\re{wtbeta}), we obtain \be \int_{\CM} A_- \wedge R df + A_+ \wedge L
df + \frac{RL}{\wt \b(X^3)} A_- \wedge A_+ \pl{IB2} \ee to be added to
\re{PS}). Comparing this with Eqs.\ \re{PSJ},\ref{Jpar}), we see that
the first two terms in \re{IB2}) give rise to a current $J^i$ which is
precisely of the form \re{Jpar}). Here this current is even simpler:
the scalar field $f$ plays the role of the two phases $\rho$ and
$\lambda$, which are {\em equal}\/ (while $r^2$ corresponds to
$R$ and $l^2$ to $L$).  The third and remaining term in \re{IB2}), on
the other hand, implies a complication: it mixes $R$- and $L$-fields
and in this respect resembles a (dilaton dependent) mass term for the
fermions.

In principle one can absorb this last term into the Poisson structure;
it still satisfies the Jacobi identity \re{Jakobi}) for $i=1,2,3$,
with $R$ and $L$ entering $\CP$ as parameters then. However, when
changing to CD-coordinates of this Poisson tensor, fields such as $\wt
A_i$ depend  implicitly on $R$ and $L$ also and thus may not be varied
independently of the latter. We therefore will not follow this route. 

Inspired by the preceding section, we want to look for chiral
solutions, i.e.\ for solutions satisfying \be d f = \ast d f\quad \LRA
\quad L = 0 \,\, , \pl{chiral} \ee where the Hodge dual is taken with
respect to $\gtrue$ (or $g$, this makes no difference for one-forms in
two dimensions). Similarly we could proceed with $ d f = -\ast d f\,\,
\LRA \,\,R=0 $. In the presence of a (Dirac) mass term, there are no
chiral fermion solutions. Similarly here for $\b'(\Phi) \not \equiv 0$
there are no solutions with $L \equiv 0$ or $R \equiv 0$. However, for
a minimal coupling, $\b= const \LRA \wt \b = const$, non--trivial
(anti--)self--dual scalar field solutions do exist.

In the preceding section we could implement a condition of the type
\re{chiral}) directly into the Lagrangian. This is no more possible
here. Still, if we keep the variation with respect to $L$, i.e.  \be
\(\wt \b(X^3) df - R A_-\) \wedge A_+ \pl{DA.4.28mod}\, , \ee as an
extra condition, we may thereafter put $L$ to zero in the Lagrangian.
But then, up to the renaming $r^2 \to R$ and $\rho \to f$, the
resulting action is {\em identical}\/ to the chiral fermion action of
the preceding section (i.e.\ to Eqs.\ \re{PSJ}, \ref{Jpar})) with
$l\equiv 0$).  Thus, with the above renaming, we may now just take the
solution of the previous section, without the need of solving any
field equation! This solution, however, has to satisfy the additional
constraint \re{DA.4.28mod}) now.  Using \re{Umrechnung}) and taking
into account that $R$ and $f$ are functions of $x^1$ only, Eq.\ 
\re{DA.4.28mod}) becomes \be R(x^1) = \wt \b(x^0) \, f'(x ^1) \, . \ee
Obviously this condition can be satisfied only for a constant function
$\wt \b$, proving our previous claim. In view of the qualitative
changes in the Klein--Gordon equation induced by a dilaton dependent
coupling, cf.\ Eq.\ \re{KG}), this result is not unplausible though
(note $df = \pm \ast df \Ra \Box f = 0$).

For $\b= const$, on the other hand, we find that, in contrast to chiral
fermions, the function $R$ may not be chosen freely, but is determined
by $f$. So the local solutions are parametrized by the choice of the
``initial data'' $f(x^1)$ and $c$. Noting, furthermore, that
${\bf T}^{ferm}_{11}=r^2 \rho'$ translates into $R f'=\b (f')^2$, which is
nothing but ${\bf T}^{scal}_{11}$ (cf.\ Eq.\ \re{Tscal}), using 
$\gtrue^{11}=0$), we arrive at the following form of the
self--dual solutions of $I_{gdil}+I_{scal}$: \be \Phi = U^{-1}(x^0) \,
, \quad f = f(x^1) \, , \ee while the metric $\gtrue$ takes the
form \re{gwahr}) with ${\bf T}_{11}={\bf T}_{11}(x^1)$ being now the only
non--vanishing component of the energy momentum tensor of the scalar
field.

Similarly the generalization of the results of the previous section 
to non--zero torsion are valid here is well. 

Concluding we remark that although for $\b=const$ the {\em general}\/
solution of the Klein--Gordon equation \re{KG}) consists of a
superposition of left-- and right--movers, $f=f_+(x^+) + f_-(x^-)$,
which are the self--dual parts of $f$, due to the non--linearities of
the coupled system of equations (cf., e.g., Eqs.\ \re{eomg2}) and
\re{Tscal})), we cannot obtain the general solution from a
superposition of the self--dual and anti--self--dual solutions found
in the present section. (A similar statement holds for fermions with
both chiralities and the chiral solutions found in the preceding
section).  The simultanous presence of chiral fermions and self--dual
scalar fields, on the other hand, is no problem; this just implies
${\bf T}_{11}={\bf T}_{11}^{ferm}+{\bf T}_{11}^{scal}$ in Eq.\ 
\re{gwahr}).

\section*{Acknowledgement}
We are grateful to M.\ Cavagli\`a, H.\ Kastrup, W.\ Kummer and A.\ 
Mundt for remarks concerning the manuscript. T.S.\ is grateful to W.\ 
Kummer for drawing his attention to the subject of chiral fermions in
connection with 2d gravitational models.

\end{document}